\documentclass[12pt,preprint]{aastex}
\usepackage{epsfig}

\def\be{\begin{equation}}
\def\ee{\end{equation}}

\def\ergs{{\rm\,erg\,s^{-1}}}

\catcode`\@=11 
\def\@versim#1#2{\vcenter{\offinterlineskip
	\ialign{$\m@th#1\hfil##\hfil$\crcr#2\crcr\sim\crcr } }}
\def\lsim{\mathrel{\mathpalette\@versim<}}
\def\gsim{\mathrel{\mathpalette\@versim>}}

\begin{document}

\title{X-ray Spectral Variability of TeV Blazars during Rapid Flares}
\author{Yongquan Xue\altaffilmark{1}, Feng Yuan\altaffilmark{2}, and Wei 
Cui\altaffilmark{1}} 
\affil{Department of Physics, Purdue University, West Lafayette, IN 47907}
\altaffiltext{1}{Email: xuey@physics.purdue.edu; cui@physics.purdue.edu}
\altaffiltext{2}{Current address: Shanghai Astronomical Observatory,
80 Nandan Road, Shanghai 200030, P.R. China; fyuan@shao.ac.cn}

\begin{abstract}

The spectral energy distribution (SED) of TeV blazars peaks both at keV 
and TeV energies. The X-ray emission is generally believed to originate 
in the synchrotron emission from relativistic electrons (and positrons) 
in the jet of these sources, while the origin of the gamma-ray emission 
is still being debated. We report results from a systematic study of X-ray 
spectral variability of Mrk 421 and Mrk 501 during individual flares 
that last for several days, making use of some of the high-quality data 
that have recently become available. The X-ray spectra of the two sources 
fall on the opposite sides of the synchrotron peak of their respective 
SEDs, so they together may offer additional insights into the physical 
origin of X-ray variability. We modeled each of the time-resolved X-ray 
spectra over a {\em single} flare by adopting a homogeneous spatial 
distribution and an instantaneous power-law spectral distribution for the 
emitting particles. We focused on the variation of four key parameters: 
particle spectral index, maximum Lorentz factor, energy density, and 
magnetic field. While there is considerable degeneracy in the fits, we 
show that, in order to account for the X-ray spectral variability observed 
in Mrk 421, at least three of the parameters are required to vary in most 
cases, with the spectral index being one of them. The observations of 
Mrk 501 support the conclusion, although the quality of the data is not 
as good. We discuss the implications of the results.

\end{abstract}


\section{Introduction}

Blazars are known to be strong gamma-ray emitters. At TeV energies, BL Lac
objects are the only type of blazars that have been detected so far. The 
spectral energy distribution (SED) of these TeV blazars invariably shows 
two characteristic peaks in the $\nu F_{\nu}$ representation, with one 
located at X-ray energies and the other at TeV energies (e.g., Fossati et 
al. 1998). While the origin of the TeV emission is still not entirely clear, 
there is a general consensus that the X-ray emission is associated with the 
synchrotron radiation from highly relativistic electrons (and positrons) in 
the jet of 
blazars. Therefore, studying the X-ray properties of TeV blazars may, in
a relatively model-independent manner, shed significant light on the 
properties of the emitting particles as well as of the environment that 
they are in.

Blazars are highly variable at nearly all wavelengths. The variability is
often in the form of discrete flares. Dramatic rapid flares have been 
observed of TeV blazars at X-ray energies (e.g., Cui 2004), as well as at 
TeV energies (e.g., Gaidos et al. 1996), with the former almost certainly 
being associated with the most energetic electrons in the jet. The observed 
X-ray flares span a wide range of timescales from many months down to less 
than an hour, and seem to form a well-organized hierarchy with weaker, 
shorter-duration flares superimposed on a stronger, longer-duration one 
(Cui 2004), perhaps indicating a scale-invariant physical origin (see also 
Xue \& Cui 2005). The flaring phenomenon could be related to internal 
shocks in the jet (Rees 1978; Spada et al. 2001), or to the ejection of 
relativistic plasma into the jet (e.g., B\"ottcher et al. 1997; 
Mastichiadis \& Kirk 1997), or to reconnection events in a magnetically 
dominated jet (Lyutikov 2003). 

While it is still difficult to pinpoint the exact mechanism at work, it may 
have become possible to investigate, quantitatively, which physical 
parameters evolve most strongly during a {\em single} flare on a timescale
of several days, thanks to improvement in the quality of X-ray data in 
recent years. In this paper, we report results 
from such an investigation with two TeV blazars, Mrk 421 and Mrk 501, which 
are among the first sources discovered at TeV energies (Punch et al. 1992; 
Quinn et al. 1996; Bradbury et al. 1997). The sources were chosen for this 
work not only because of the availability of high-quality data but also of 
their diverse behaviors at X-ray energies. In particular, the observed X-ray 
spectra of the two sources fall on the opposite sides of the synchrotron 
peak in their respective SEDs. Modeling their spectra together may, therefore,
help reveal common threads underlying the flaring phenomenon in TeV blazars.

\section{Observational Aspects}

\subsection{Data}

We observed Mrk 421 regularly with {\em RXTE} in 2003 and 2004, as part of a 
multiwavelength monitoring campaign (B{\l}a\.zejowski et al. 2005). The source
became exceptionally bright in X-rays near the end of the campaign. Figure~1 
shows the relevant portion of the (daily-averaged) ASM light curve. A number 
of distinct flares are present, with the strongest one peaking roughly at 
about 10 c/s (or 135 mcrab in the ASM band of 1.5--12 keV)! For detailed 
spectral 
studies, we used data from the pointed {\em RXTE} observations taken with 
the Proportional Counter Array (PCA). The times of these observations are 
noted in Fig.~1. The PCA detector consists of five nearly identical 
proportional counter units (PCUs). However, only PCU 0 and PCU 2 were in use 
throughout the campaign. PCU 0 has lost its front veto layer, so the data 
from it are prone to contamination by events caused by low-energy electrons 
entering the detector. For this work, therefore, we chose PCU 2 as our 
``standard'' detector for flux normalization and spectral analysis. We also 
only used data taken in the $Standard2$-mode, which have a time resolution 
of 16 seconds and cover the full energy range (nominally 2--60 keV). 

We followed Cui (2004) closely in reducing and analyzing the PCA data. 
Briefly, the data were reduced with {\em ftools v5.2}. For a given
observation, we first filtered data by following the standard procedure for 
faint sources,\footnote{See the online {\em RXTE} Cook Book at 
http://heasarc.gsfc.nasa.gov/docs/xte/recipes/cook\_book.html.} which 
resulted in a list of good time intervals (GTIs). We then simulated
background events for the observation by using the latest background model 
(pca\_bkgd\_cmfaintl7\_eMv20031123.mdl) that is appropriate for faint 
sources. Using the GTIs, we proceeded to 
make a spectrum for each PCU by using data from only the first xenon layer 
(which is most accurately calibrated), which limits the spectral coverage 
to roughly 3--25 keV. Since few counts are detected at higher energies, 
the impact of the reduced spectral coverage is very minimal. We repeated 
the steps to derive a corresponding background spectrum for the PCU from 
the simulated events.

Unlike Mrk 421, Mrk 501 has been fairly quiet in recent years. For this work, 
we obtained archival {\em RXTE} data taken during the 1997 outburst of the
source. For direct comparisons we treated the data in the exact same manner 
as in the case of Mrk 421.

\subsection{Results}

Figure 2 shows the PCA light curves of Mrk 421 and Mrk 501, covering the
respective periods of strong flares. The coverage gaps in both cases are 
mainly caused by the requirement for coordination with ground-based facilities.
For this work, we would ideally like to focus on one single flare and follow
the spectral evolution of the source through both the rising and decaying 
periods. We were able to do so in the case of Mrk 421. A large flare was well 
sampled by the observations, as shown in Fig.~2, and we chose five of the 
observations for further analyses. As pointed out by Cui (2004), however, 
it is, strictly speaking, not possible to isolate a flare from the 
flaring hierarchy. The notion of ``single'' flares is therefore meaningful 
only when a timescale is specified. Here, we are specifically
interested in flares that last for days. In the case of Mrk 501, no single 
flare was adequately sampled. We selected representative observations that 
covered the rising phase of one flare and the decaying phase of a subsequent 
flare, as also shown in Fig.~2. Note that Mrk 501 was much fainter than 
Mrk 421.

For each observation, we jointly fitted the PCU spectra in {\em XSPEC v11.3.1} 
(Arnaud 1996). An additional multiplicative factor was introduced to take 
into account any uncalibrated difference in the overall throughput of the
PCUs, with reference to PCU 2. Following the usual practice, we added 1\% 
systematic uncertainty to the data. Given the lack of sensitivity of the
PCA below about 2 keV, we adopted the Galactic values for the line-of-sight
hydrogen column densities ($1.38$ and $1.8\times 10^{20}$ cm$^{-2}$ for 
Mrk~421 and Mrk~501, respectively; Dickey \& Lockman 1990) and fixed them 
in the fitting process.

We experimented with three empirical models: power law, broken power law, 
and power law with an exponential cut-off. The cut-off power law provides 
best fits to the data for the selected observations of Mrk 421 (as shown 
in Fig.~3), although fits with the broken power law are also acceptable.
For Mrk 501, the broken and cut-off power law models fit the data almost 
equally well. In both cases, we used the best-fit model to unfold each 
observed (count) spectrum to derive the corresponding SED for further 
theoretical modeling. We should note that only the PCU2 data were used 
to produce the SED, while data from all active PCUs were used to find
the best-fit model. 

Figure~3 shows representative X-ray SEDs for Mrk 421 and Mrk 501, 
respectively. Both sources exhibit significant spectral variability during 
the flares. In general, the spectrum seems to harden toward high fluxes. It
is also clear, from the figure, that the measured X-ray SEDs of Mrk 421 
and Mrk 501 lie on the opposite sides of their respective (synchrotron) 
peaks, which are (mostly) below 2 keV in the former case but above 20 keV 
in the latter. Therefore, together, the sources should be representative 
of TeV blazars. 

\section{Theoretical Aspects}

To gain insights into variability of the underlying distribution of electrons,
as well as their environment, we proceeded to construct a physical model and 
to apply the model to the data in a systematic manner.

\subsection{Model}

We assume that each flare of interest originates in one localized region 
in the jet and that the electron and magnetic field are distributed
homogeneously in the region. We further assume that the energy spectrum 
of {\em emitting} electrons is of power-law shape, 
\be
n(\gamma)d\gamma=N_0\gamma^{-p}d\gamma, \hspace{1cm}
\gamma_{\rm min}\le\gamma\le\gamma_{\rm max}.
\ee
The total energy density of electrons is then
\be
E_{\rm tot}=N_0m_ec^2\frac{\gamma_{\rm min}^{2-p}-\gamma_{\rm max}^{2-p}}{p-2}, 
\hspace{1cm} p\ne 2.
\ee

For an electron with Lorentz factor of $\gamma$, the differential power of its 
synchrotron radiation is given by
\be
\frac{dP(\nu)}{d\nu}=\frac{2\pi\sqrt{3}e^2\nu_L}{c}\left[\frac{\nu}{\nu_c}\int^{\infty}_{\nu/\nu_c}K_{5/3}(t)dt\right](\ergs~{\rm Hz}^{-1}),
\ee
where the Larmor frequency $\nu_L=\frac{1}{2\pi}\frac{eB}{m_ec}$, the critical
frequency $\nu_c=\frac{3}{2}\nu_L\gamma^2$, and $K_{n}(t)$ is the modified 
Bessel function of the second kind of order $n$. Assuming that the flaring
region is of spherical shape with radius $r$, the spectrum of the integrated
emission (in the jet frame) is
\be\label{Fnu}
F_{\nu}=\frac{4}{3}\pi r^3\int^{\gamma_{\rm max}}_{\gamma_{\rm min}}n(\gamma)\frac{dP(\nu)}{d\nu}d\gamma.
\ee
If the Doppler factor of the jet is $\delta$, the observed spectrum is given
by
\be
F_{\rm obs}(\nu^{\rm obs})=\delta^3~F_{\nu}(\nu^{\rm obs}/\delta)
\ee
It is often taken casually that such an optically-thin synchrotron spectrum 
is also of power-law shape with a spectral index of $\alpha=(p-1)/2$. This 
is only true when the portion of the spectrum of interest is far away from 
the critical frequencies corresponding to either $\gamma_{\rm min}$ or 
$\gamma_{\rm max}$. The issue is of particular relevance to the case of 
Mrk~421 because its X-ray spectrum seems to be entirely associated with 
electrons near $\gamma_{\rm max}$, as shown in Fig.~3. Therefore, care must 
be taken in the calculation.

There are many parameters in the model. The objective of this work is to 
systematically investigate the roles of ``key'' parameters in producing the 
observed X-ray spectral variability during a major flare. We identified 
these parameters as: $p$, $\gamma_{\rm max}$, $E_{\rm tot}$, and $B$. We 
adopted $E_{\rm tot}$ rather than $N_0$ because (1) it seems obvious that
$N_0$ must vary during a flare, so few insights would be gained; and
(2) $E_{\rm tot}$, though depending on other parameters of interest, is a 
well-defined physical quantity and provide important information on the 
energetics of the flaring process. All other parameters were fixed during 
the fitting process. 
The size of the emitting region was chosen to be compatible with the observed 
timescale of the flare, $r=1\times 10^{16}$ and $5\times 10^{15}$ cm for 
Mrk~421 and Mrk~501, respectively; the Doppler factor was set at a nominal 
value for TeV blazars ($\delta=15$); and the minimum Lorentz factor of the 
electrons was made sufficiently low ($\gamma_{\rm min}=10^4$) such that it 
does not affect the conclusions reached. For this work, we adopted the 
luminosity distances of $d=137$ and $156$ Mpc for Mrk~421 ($z=0.030$) and 
Mrk~501 ($z=0.034$), respectively, which were derived with the following 
cosmological parameters: 
$H=65$ km s$^{-1}$ Mpc$^{-1}$, $\Omega_m=0.3$, and $\Omega_{\Lambda}=0.7$.

\subsection{Results}

For each spectrum shown in Fig.~3, a grid search was carried out to find
statistically acceptable solutions. A solution is defined as a combination 
of $p$, $\gamma_{\rm max}$, $E_{\rm tot}$, and $B$ that leads to a fit with
reduced $\chi^2$ in the range of $1\pm \sqrt{2/\nu}$, where $\nu$ is the 
degree of freedom of the fit. Note that a $\chi^2$ distribution can be 
approximated by a normal distribution with mean of $\nu$ and standard 
deviation of $\sqrt{2\nu}$ when $\nu$ is sufficiently large. Here, $\nu$
is typically $30$--$40$. The ranges of the parameters covered in the search 
are: $p$ in 1.00--4.00, $B$ in $10^{-3}$--$10^2$ G, $\gamma_{\rm max}$ in 
$10^5$--$10^{11}$, and $E_{\rm tot}$ in $10^8$--$10^{15}$ cm$^{-3}$. 
The step used is linear for $p$, with $\Delta p = 0.01$, but logarithmic 
for $B$, $\gamma_{\rm max}$, and $E_{\rm tot}$, with 50 and 25 points per 
decade for Mrk~421 and Mrk~501, respectively. 

Table~1 summarizes the results for Mrk~421 and Mrk~501 separately. The ranges 
shown indicate the portion of the parameter space where solutions are found. 
It is quite clear that $p$ is quite well constrained in both cases. For
illustration, Fig.~4 shows the distributions of $p$ for Mrk~421. The $p$ 
distribution is nicely peaked. The most probable values of $p$ are also 
shown in Table~1 for both sources. The results seems to indicate that the 
electron spectrum tends to harden toward the peak of a flare. There is a 
substantial amount of degeneracy among the other three parameters: 
$\gamma_{\rm max}$ is fairly well constrained in Mrk~421 but not at all in 
Mrk~501 (presumably due to poorer data quality); $B$ is poorly constrained 
in both cases; and $E_{\rm tot}$ is not constrained at all. The distributions 
of these parameters are fairly flat with no dominant peaks. The best fits are 
shown in Fig.~3.

To break the degeneracy, we proceeded to examine specialized cases, in which 
a subset of the parameters are fixed at some ``typical'' values, as seen 
from modeling broadband SEDs of the sources. It became immediately clear 
that, to account for the observed X-ray spectral variability of Mrk~421, $p$ 
{\em cannot} be fixed. This is not the case for Mrk~501, presumably due to
the poorer statistics again. Taking the view that the same processes are 
operating in both sources, we allowed $p$ to vary in all subsequent analyses.

We first experimented with fixing one of the remaining parameters, $B$, 
$\gamma_{\rm max}$, and $E_{\rm tot}$. We found acceptable fits to {\em all} 
of the spectra in all cases. As examples, Table~2 and Table~3 show the results 
obtained with $B=0.10$, $0.83$ G and $\gamma_{\rm max}=5.0\times 10^5$, 
$5.2\times 10^6$ for Mrk~421 and Mrk~501, respectively (see, e.g., 
B{\l}a\.zejowski et al 2005 and Pian et al. 1998). The free parameters are 
now much more constrained. We then proceeded to investigate whether we could
fix two of the parameters and still obtain solutions. We found that it was
not possible to account for the observed spectral variability in Mrk~421, 
except for the case in which $B$ and $\gamma_{\rm max}$ were fixed, which 
confirms that the synchrotron peak of the SED ($\propto \gamma_{\rm max}^2 B$) 
did not vary too much during the flare. As for Mrk~501, the data are 
much less constraining; we were still able to find satisfactory fits to 
the spectra. For completeness, we also examined the possibility of fixing 
all three parameters. In this case, we failed to obtain acceptable fits to 
{\em all} of the spectra for either source.
 
It is often assumed that the ratio of the energy of the radiating particles
to that of the magnetic field remains constant, in theoretical studies of 
gamma ray bursts and blazars. To check if the assumption can be accomodated
by the data, we looked at cases in which $E_{\rm tot}/B^2$ remains constant.
For both sources, acceptable solutions were found. As an example, we show
in Table~4 those solutions selected from the full set (as shown in Table~1)
that have $E_{\rm tot}/B^2$ fixed at some fiducial values. In this case, 
the other parameters are even more constrained, although it is not possible
to know the exact value of the ratio {\em a priori}.

\section{Discussion}

We have obtained time-resolved X-ray spectra of Mrk~421 during the rise and 
fall of a prominent X-ray flare, as well as of Mrk~501 during two consecutive 
flares (on a timescale of a few days) for comparison. In general, the X-ray 
spectrum steepens as the flux 
decreases (see Fig.~3), as was seen previously in X-ray selected BL Lac 
objects in general (Giommi et al. 1990). For Mrk~421, the synchrotron peak
is clearly below the low-energy sensitivity threshold of the PCA detector 
($\sim$2 keV), judging from the shape of the SEDs, except for one case 
(``$t_3$'' in Fig.~3) in which the peak appears to have moved into the PCA 
passing band. A shift in the synchrotron peak with flux has been seen in 
Mrk~501 previously (Pian et al. 1998; Sambruna et al. 2000). In our case, 
the synchrotron peak of Mrk~501 is clearly above the spectral range of the 
data shown in Fig.~3 (but not necessarily above that of the PCA), which is 
known (e.g., Pian et al. 1998; Catanese et al. 1997; Lamer \& Wagner 1998), 
and is also why we selected it for comparison with Mrk~421 in the first place.
Because of that, our SED modeling suffers from serious degeneracy between $B$ 
and $\gamma_{\rm max}$. For instance, the unusually-high values allowed for 
$\gamma_{\rm max}$ or unusually-low values for $B$ is the consequence of 
such uncertainty.

To cast light on the origin of the observed X-ray spectral variability of
both Mrk~421 and Mrk~501, we took an approach to model {\em instantaneous}
properties of the sources throughout the flares. In other words, we were 
{\em not} concerned with the steady-state properties. For simplicity, we 
assumed that the emitting electrons always follow a power-law spectral 
distribution, which may or may not be the case in reality, after heating
and cooling processes are properly taken into account. Moreover, we 
assumed that the spatial distribution of the electrons is homogeneous 
during a flare. Even with such an idealistic approach, the model still 
contains too many parameters to de-couple completely. We narrowed our focus 
onto the evolution of four key parameters: $p$, $\gamma_{\rm max}$, 
$E_{\rm tot}$, and $B$, with other parameters fixed at nominal values 
derived from variability timescales and broadband SEDs (in particular, the 
TeV gamma-ray properties). In spite of serious degeneracy in the solutions, 
we are able to reach a few conclusions that we think are robust.    

As expected, the high-quality data of Mrk~421 have provided the most severe
constraints on the model. To account for the observed spectral variability
of the source, we found that the electron spectral index ($p$) must vary. 
There appears to be a general trend of decreasing $p$ toward high fluxes. 
The only exception seen ($t_1$ in Fig.~4) is probably due to ``contamination''
by a previous flare (see Fig.~2). Of course, it could also be associated 
with local variability {\em within} the flare of interest. Again, there may 
always be smaller flares superimposed on it (Cui 2004). The change in $p$ 
alone is, however, 
not sufficient for explaining the observed X-ray spectral variability. At 
least two of the three remaining parameters must also vary, in all but one 
combinations. The lone exception is that acceptable solutions do exist even 
when $B$ and $\gamma_{\rm max}$ are both fixed. This is an indication that 
the synchrotron peak did not vary too much during the flare. Even when all 
four parameters are allowed to vary, both $p$ and $\gamma_{\rm max}$ are 
reasonably well constrained (see Table~1). If we set $B$ or $\gamma_{\rm max}$
at typical values from modeling the broadband SED, as well as time 
variability, in the context of the synchrotron self-Compton scenario, the 
other parameters are much more constrained and their values are also quite 
typical (see Tables 2 and 3).

The quality of data is not nearly as good for Mrk~501 as for Mrk~421. 
However, we feel that it is a worthwhile exercise to see whether the same 
model can also be applied to a different TeV blazar, like Mrk~501 whose 
observed X-ray spectrum is almost entirely below the synchrotron peak. We 
followed precisely the same procedures in modeling the X-ray spectra of 
Mrk~501 as Mrk~421. From Tables 1, 2, and 3, it is clear that the model 
parameters are poorly constrained in this case. One interesting (and solid)
result is that the electron spectrum is much harder for Mrk~501 than for
Mrk~421, even after taking into account the large uncertainties. Moreover,
the electron spectrum hardens toward high fluxes during the rising phase 
of the first flare, although the trend is not as clear during the decaying
phase of the second flare, perhaps due to the presence of more complicated
local variability (see Fig.~2). Because the synchrotron peak is beyond the 
energy range covered by the data in this case, $B$ and $\gamma_{\rm max}$ 
are hardly constrained. 

We should stress that it is not our goal to exhaustively explore the 
entire parameter space of the model. The scope of our investigation is
limited only to a few selected parameters, which could affect the 
conclusions. First of all, although the effects of some 
of the parameters not selected (e.g., the size of the emitting region) 
can be absorbed into $E_{\rm tot}$, those of others may still contribute 
to the observed X-ray spectral variability. For instance, the Doppler 
factor of the jet might not remain constant during a flare and thus lead 
to not only variable Doppler boosting of the source intensity (which only
affects the overall normalization) but also variable Doppler shift of the 
photon energy (see Eq. 5). We should note, however, that we also ran a 
case for Mrk 421 where 
the size of the emitting region and the Doppler factor were fixed at very 
different values ($R=10^{15}$ cm and $\delta = 50$). Although the exact
values of the best-fit parameters are different, our conclusions remain
the same. Second, the homogeneous electron distribution considered 
here may be a gross over-simplification of the reality. Multiple electron
populations may be present, with different properties and environments, 
and all may contribute to the X-ray emission. However, in our view, it 
seems more likely that a single rapid flare is associated with one electron 
population, which is why we tried to focus on specific flares. The most 
important conclusion of this study is that the observed X-ray spectral
variability of TeV blazars is likely caused by a combination of spectral 
variation of the radiating electrons and changes in the environment 
(e.g., $B$). This is hardly surprising because it seems unlikely that, 
in reality, the triggering of a flare involves changes only in one of the 
parameters. Moreover, our results show that the spectral index ($p$) of
the particle distribution must vary during a flare, together with the 
overall normalization ($N_0$).

\acknowledgments
This work was supported in part by NASA Grants NAG5-13736 and NNG04GP86G.
It has made use of data obtained through the High Energy Astrophysics 
Science Archive Research Center Online Service, provided by the NASA/Goddard 
Space Flight Center, and through the NASA/IPAC Extragalactic Database, 
which is operated by the Jet Propulsion Laboratory, California Institute 
of Technology. 

\clearpage

\begin{table}
\caption{Spectral Solutions: Full Set}
\begin{tabular}{cccccc}\hline\hline
Observation & MJD & $p$ & $B$ & $\gamma_{\rm max}$  & $E_{\rm tot}$ \\
(Data Set) &  &   & (Gauss) &  & (cm$^{-3}$)$^a$ \\ \hline
& &  mpv(min,max)$^b$ & (min,max) & (min,max) & (min,max) \\ \hline
421 $t1$ & 53114.25 & 3.16(3.04,3.27) & (0.002,3.98) & (1.9$\times 10^5$,8.3$\times 10^6$) & (10$^8$,10$^{15}$) \\
421 $t2$ & 53115.31 & 3.45(3.20,3.64) & (0.003,5.25) & (1.5$\times 10^5$,6.0$\times 10^6$) & (10$^8$,10$^{15}$) \\
421 $t3$ & 53116.30 & 2.76(2.54,2.93) & (0.002,4.57) & (1.7$\times 10^5$,9.5$\times 10^6$) & (10$^8$,10$^{15}$) \\
421 $t4$ & 53118.24 & 3.39(3.13,3.57) & (0.003,5.50) & (1.4$\times 10^5$,6.0$\times 10^6$) & (10$^8$,10$^{15}$) \\
421 $t5$ & 53119.25 & 3.87(3.26,3.95) & (0.002,4.37) & (1.7$\times 10^5$,1.0$\times10^{11}$) & (10$^8$,10$^{15}$) \\\hline
501 $T1$ & 50553.37 & 2.60(1.77,2.73) & (0.002,75.9) & (1.7$\times 10^5$,1.0$\times10^{11}$) & (10$^8$,10$^{15}$) \\
501 $T2$ & 50554.17 & 2.28(1.35,2.46) & (0.002,100.) & (1.7$\times 10^5$,1.0$\times10^{11}$) & (10$^8$,10$^{15}$) \\
501 $T3$ & 50554.44 & 2.13(1.04,2.45) & (0.002,100.) & (1.6$\times 10^5$,1.0$\times10^{11}$) & (10$^8$,10$^{15}$) \\
501 $T4$ & 50580.36 & 2.56(1.76,2.71) & (0.002,91.2) & (1.7$\times 10^5$,1.0$\times10^{11}$) & (10$^8$,10$^{15}$) \\
501 $T5$ & 50581.19 & 2.30(1.39,2.71) & (0.002,83.2) & (1.6$\times 10^5$,1.0$\times10^{11}$) & (10$^8$,10$^{15}$) \\
501 $T6$ & 50581.36 & 2.07(1.57,2.37) & (0.002,6.31) & (1.9$\times 10^5$,1.3$\times 10^{7}$) & (10$^8$,10$^{15}$) \\
501 $T7$ & 50582.36 & 2.63(1.32,2.72) & (0.002,75.9) & (1.6$\times 10^5$,1.0$\times10^{11}$) & (10$^8$,10$^{15}$) \\\hline
\end{tabular}
\begin{flushleft}
$^a$ $E_{tot}$ has been scaled by the electron rest energy \\
$^b$ mpv: most probable value
\end{flushleft}
\end{table}

\begin{table}
\caption{Spectral Solutions: Subset with Fixed $B$}
\begin{tabular}{cccccc}\hline\hline
Observation & MJD & $p$ & $B$ & $\gamma_{\rm max}$  & $E_{\rm tot}$ \\
(Data Set) &  &   & (Gauss) &  & (cm$^{-3}$) \\ \hline
& &  mpv(min,max) & (fixed) & (min,max) & (min,max) \\ \hline
421 $t1$ & 53114.25 & 3.11(3.05,3.23) & 0.10 & (1.1$\times 10^6$,1.3$\times 10^6$) & (2.5$\times 10^{11}$,3.5$\times 10^{11}$) \\
421 $t2$ & 53115.31 & 3.48(3.22,3.62) & 0.10 & (1.0$\times 10^6$,1.3$\times 10^6$) & (4.6$\times 10^{11}$,1.1$\times 10^{12}$) \\
421 $t3$ & 53116.30 & 2.75(2.62,2.89) & 0.10 & (1.1$\times 10^6$,1.3$\times 10^6$) & (2.3$\times 10^{11}$,3.2$\times 10^{11}$) \\
421 $t4$ & 53118.24 & 3.34(3.14,3.56) & 0.10 & (1.0$\times 10^6$,1.3$\times 10^6$) & (4.6$\times 10^{11}$,1.1$\times 10^{12}$) \\
421 $t5$ & 53119.25 & 3.88(3.30,3.95) & 0.10 & (9.5$\times 10^5$,1.0$\times 10^{11}$) & (2.5$\times 10^{11}$,1.2$\times 10^{12}$) \\\hline
501 $T1$ & 50553.37 & 2.60(1.91,2.71) & 0.83 & (4.8$\times 10^5$,1.0$\times 10^{11}$) & (4.8$\times 10^9$,2.3$\times 10^{12}$) \\
501 $T2$ & 50554.17 & 2.28(1.75,2.34) & 0.83 & (6.3$\times 10^5$,1.0$\times 10^{11}$) & (8.3$\times 10^9$,7.6$\times 10^{13}$) \\
501 $T3$ & 50554.44 & 1.97(1.04,2.31) & 0.83 & (4.4$\times 10^5$,2.5$\times 10^{6}$) & (6.9$\times 10^9$,2.3$\times 10^{10}$) \\
501 $T4$ & 50580.36 & 2.56(1.94,2.63) & 0.83 & (5.2$\times 10^5$,1.0$\times 10^{11}$) & (6.3$\times 10^{9}$,2.8$\times 10^{12}$) \\
501 $T5$ & 50581.19 & 2.42(1.62,2.63) & 0.83 & (4.4$\times 10^5$,5.8$\times 10^{9}$) & (5.2$\times 10^{9}$,2.8$\times 10^{11}$) \\
501 $T6$ & 50581.36 & 2.04(1.78,2.31) & 0.83 & (4.8$\times 10^5$,6.9$\times 10^{5}$) & (5.2$\times 10^{9}$,6.9$\times 10^{9}$) \\
501 $T7$ & 50582.36 & 2.63(1.38,2.67) & 0.83 & (3.6$\times 10^5$,9.1$\times 10^{10}$) & (3.6$\times 10^{9}$,6.3$\times 10^{11}$) \\\hline
\end{tabular}
\end{table}

\begin{table}
\caption{Spectral Solutions: Subset with Fixed $\gamma_{\rm max}$}
\begin{tabular}{cccccc}\hline\hline
Observation & MJD & $p$ & $B$ & $\gamma_{\rm max}$  & $E_{\rm tot}$ \\
(Data Set) &  &   & (Gauss) &  & (cm$^{-3}$) \\ \hline
& &  mpv(min,max) & (min,max) & (fixed) & (min,max) \\ \hline
421 $t1$ & 53114.25 & 3.09(3.07,3.12) & (0.52,0.58) & 5.0$\times 10^5$ & (6.6$\times 10^{9}$,7.6$\times 10^{9}$) \\
421 $t2$ & 53115.31 & 3.45(3.23,3.60) & (0.40,0.69) & 5.0$\times 10^5$ & (1.1$\times 10^{10}$,2.5$\times 10^{10}$) \\
421 $t3$ & 53116.30 & 2.76(2.59,2.89) & (0.50,0.76) & 5.0$\times 10^5$ & (4.6$\times 10^{9}$,8.3$\times 10^{9}$) \\
421 $t4$ & 53118.24 & 3.39(3.13,3.56) & (0.40,0.69) & 5.0$\times 10^5$ & (1.3$\times 10^{10}$,2.4$\times 10^{10}$) \\
421 $t5$ & 53119.25 & 3.80(3.29,3.92) & (0.36,4.17) & 5.0$\times 10^5$ & (1.0$\times 10^{8}$,1.6$\times 10^{10}$) \\\hline
501 $T1$ & 50553.37 & 2.58(1.88,2.70) & (0.01,15.8) & 5.2$\times 10^6$ & (1.0$\times 10^8$,8.3$\times 10^{13}$) \\
501 $T2$ & 50554.17 & 2.26(1.35,2.34) & (0.01,30.2) & 5.2$\times 10^6$ & (1.0$\times 10^8$,1.0$\times 10^{14}$) \\
501 $T3$ & 50554.44 & 1.98(1.04,2.32) & (0.01,0.25) & 5.2$\times 10^6$ & (2.8$\times 10^{11}$,1.4$\times 10^{14}$) \\
501 $T4$ & 50580.36 & 2.55(1.76,2.62) & (0.01,17.4) & 5.2$\times 10^6$ & (1.0$\times 10^{8}$,1.0$\times 10^{14}$) \\
501 $T5$ & 50581.19 & 2.29(1.60,2.58) & (0.01,0.05) & 5.2$\times 10^6$ & (2.8$\times 10^{12}$,1.3$\times 10^{14}$) \\
501 $T6$ & 50581.36 & 2.04(1.58,2.36) & (0.01,0.02) & 5.2$\times 10^6$ & (2.1$\times 10^{13}$,1.2$\times 10^{14}$) \\
501 $T7$ & 50582.36 & 2.60(1.32,2.66) & (0.01,13.2) & 5.2$\times 10^6$ & (1.0$\times 10^{8}$,1.6$\times 10^{14}$) \\\hline
\end{tabular}
\end{table}

\begin{table}
\caption{Spectral Solutions: Subset with Fixed $E_{\rm tot}/B^2$}
\begin{tabular}{cccccc}\hline\hline
Observation & MJD & $p$ & $B$ & $\gamma_{\rm max}$  & $E_{\rm tot}$ \\
(Data Set) &  &   & (Gauss) &  & (cm$^{-3}$) \\ \hline
& &  mpv(min,max) & (min,max) & (min,max) & (min,max) \\ \hline
421 $t1$ & 53114.25 & 3.11(3.05,3.16) & (0.10,0.11) & (1.1$\times 10^6$,1.2$\times 10^6$) & (2.5$\times 10^{11}$,2.8$\times 10^{11}$) \\
421 $t2$ & 53115.31 & 3.49(3.31,3.59) & (0.12,0.14) & (9.5$\times 10^5$,1.1$\times 10^6$) & (3.6$\times 10^{11}$,4.8$\times 10^{11}$) \\
421 $t3$ & 53116.30 & 2.78(2.70,2.86) & (0.10,0.11) & (1.2$\times 10^6$,1.3$\times 10^6$) & (2.5$\times 10^{11}$,2.8$\times 10^{11}$) \\
421 $t4$ & 53118.24 & 3.36(3.23,3.52) & (0.12,0.14) & (9.5$\times 10^5$,1.0$\times 10^6$) & (3.6$\times 10^{11}$,4.8$\times 10^{11}$) \\
421 $t5$ & 53119.25 & 3.83(3.30,3.91) & (0.10,0.14) & (9.5$\times 10^5$,1.0$\times 10^{11}$) & (2.5$\times 10^{11}$,4.8$\times 10^{11}$) \\\hline
501 $T1$ & 50553.37 & 2.60(2.27,2.71) & (0.28,1.20) & (1.1$\times 10^6$,1.0$\times 10^{11}$) & (5.8$\times 10^{10}$,1.1$\times 10^{12}$) \\
501 $T2$ & 50554.17 & 2.28(2.15,2.34) & (0.33,3.31) & (1.9$\times 10^6$,1.0$\times 10^{11}$) & (8.3$\times 10^{10}$,8.3$\times 10^{12}$) \\
501 $T3$ & 50554.44 & 1.85(1.70,2.27) & (0.30,0.36) & (1.0$\times 10^6$,2.8$\times 10^{6}$) & (6.9$\times 10^{10}$,1.0$\times 10^{11}$) \\
501 $T4$ & 50580.36 & 2.56(1.92,2.62) & (0.28,1.20) & (9.1$\times 10^5$,1.0$\times 10^{11}$) & (5.8$\times 10^{10}$,1.1$\times 10^{12}$) \\
501 $T5$ & 50581.19 & 2.46(2.05,2.60) & (0.28,0.58) & (9.1$\times 10^5$,5.2$\times 10^{8}$) & (5.8$\times 10^{10}$,2.5$\times 10^{11}$) \\
501 $T6$ & 50581.36 & 2.18(2.15,2.22) & (0.28,0.28) & (1.0$\times 10^6$,1.0$\times 10^{6}$) & (5.8$\times 10^{10}$,5.8$\times 10^{10}$) \\
501 $T7$ & 50582.36 & 2.63(1.82,2.66) & (0.25,0.76) & (8.3$\times 10^5$,3.3$\times 10^{10}$) & (4.8$\times 10^{10}$,4.4$\times 10^{11}$) \\\hline
\end{tabular}
\end{table}

\clearpage
\begin{figure}
\epsscale{1.}
\plotone{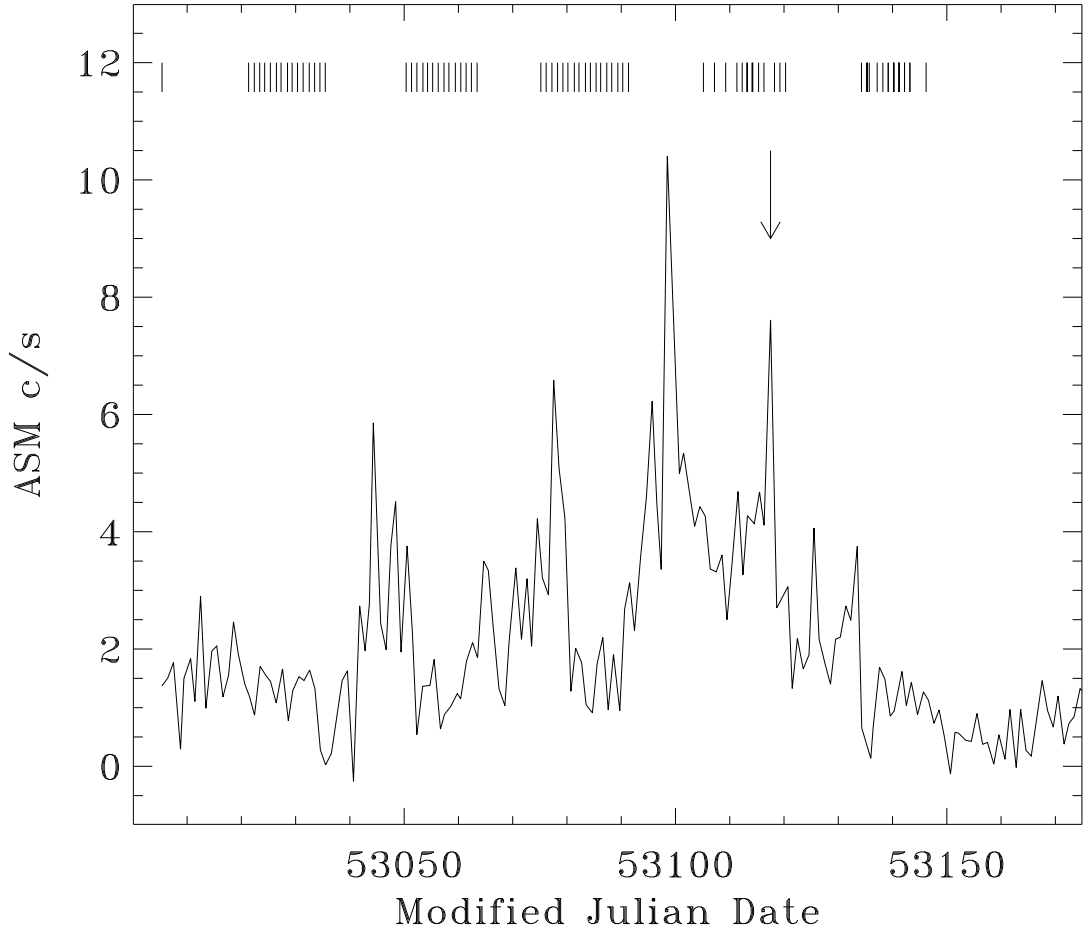}
\caption{Daily-binned ASM light curve of Mrk~421 in 2004. Each data point 
represents a weighted average of raw count rates in the summed band (1.5--12 
keV) over 1 day. For clarity, the error bars are not shown. The times of the 
pointed PCA/$RXTE$ observations are indicated with vertical lines at the top. 
The arrow points to a flare that was very well covered by the PCA
observations. For reference, the Crab Nebula produces about 75 c/s. }
\end{figure}

\begin{figure}
\epsscale{1.}
\plotone{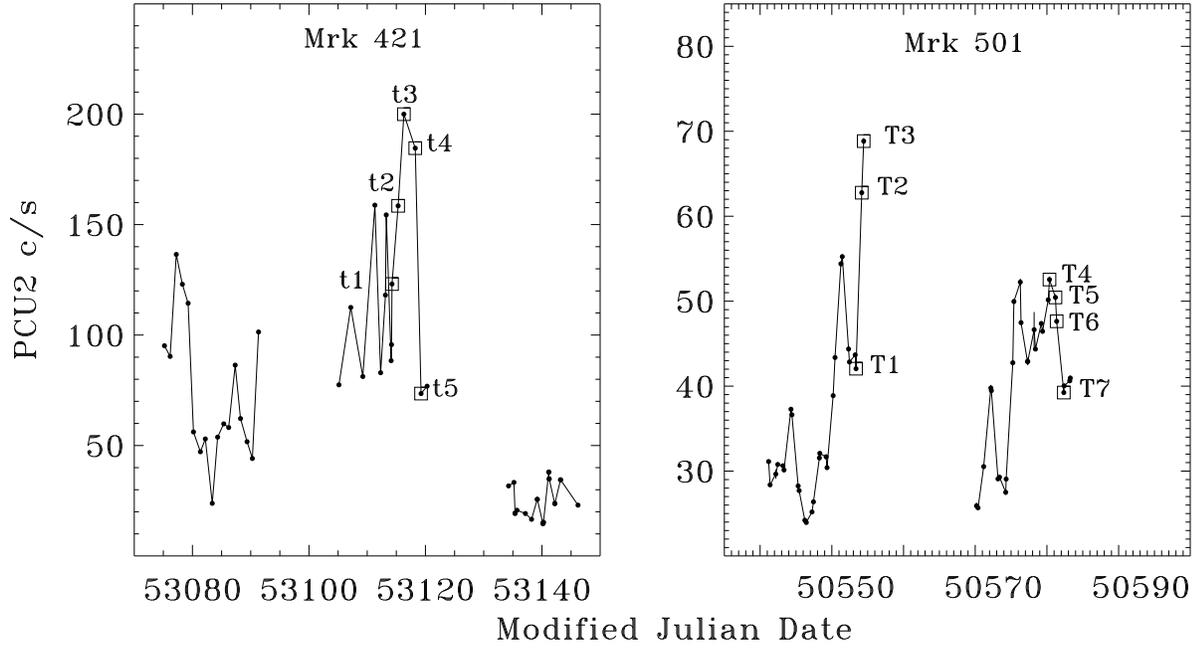}
\caption{PCA light curves of Mrk~421 in 2004 ({\em left}) and Mrk~501 in 1997
({\em right}). The results were derived from the PCU 2 data alone. The count 
rates were computed in the 2--60 keV band. For reference, the Crab Nebula
produces about 2700 c/s/PCU. The observations selected for 
subsequent analyses are marked. Note the presence of coverage gaps. }
\end{figure}

\begin{figure}
\epsscale{1.}
\plotone{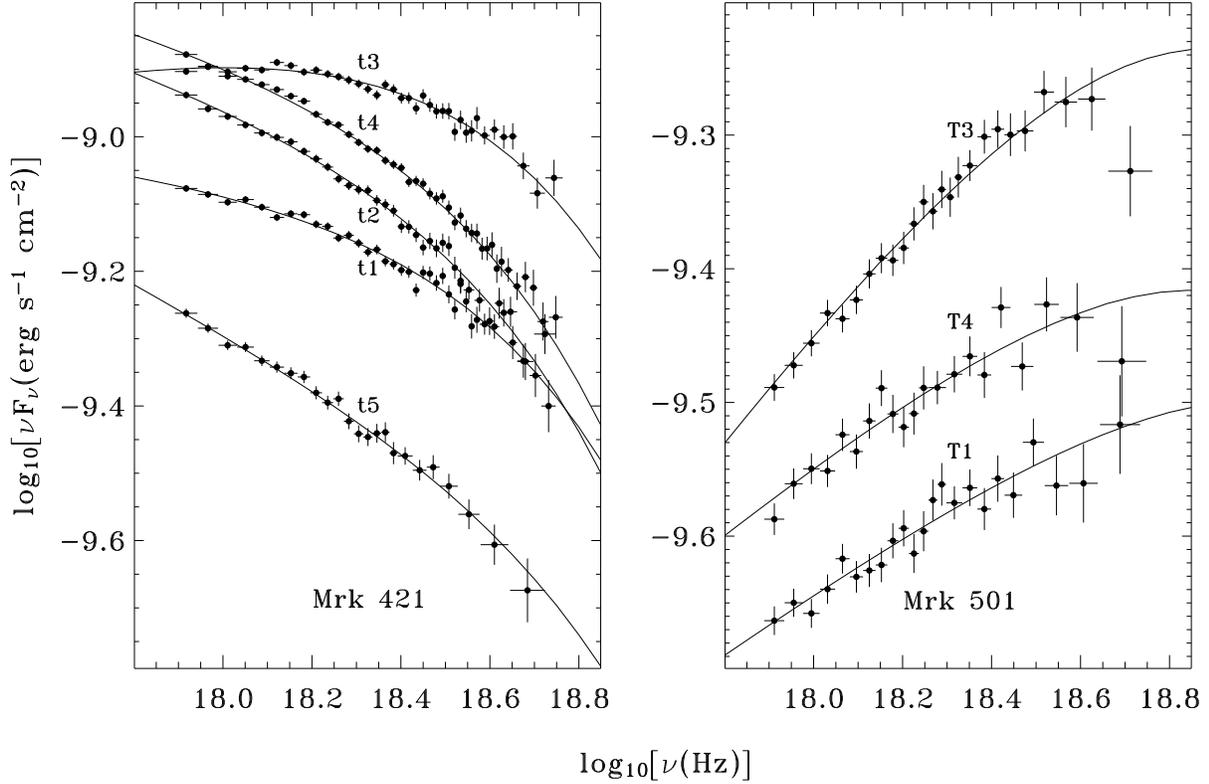}
\caption{Representative X-ray spectra of Mrk~421 ({\em left}) and Mrk~501
({\em right}). The solid lines show the best-fit models. The parameters 
are: for Mrk~421, ({\it t1}: $p=3.16$, $B=0.22$, 
$\gamma_{\rm max}=8.2\times 10^5$,
$E_{\rm tot}=1.05\times 10^{11}$), ({\it t2}: $p=3.45$, $B=0.25$, 
$\gamma_{\rm max}=7.1\times 10^5$, $E_{\rm tot}=1.75\times 10^{11}$), ({\it t3}: 
$p=2.72$, $B=0.21$, $\gamma_{\rm max}=8.3\times 10^5$, 
$E_{\rm tot}=1.05\times 10^{11}$), ({\it t4}: $p=3.38$, $B=0.31$, 
$\gamma_{\rm max}=6.3\times 10^5$, $E_{\rm tot}=1.1\times 10^{11}$), ({\it t5}: 
$p=3.71$, $B=0.29$, $\gamma_{\rm max}=7.9\times 10^5$, 
$E_{\rm tot}=0.95\times 10^{11}$); for Mrk~501, ({\it T1}: $p=2.52$, $B=0.10$, 
$\gamma_{\rm max}=3.1\times 10^6$, $E_{\rm tot}=5.0\times 10^{11}$), ({\it T3}: 
$p=2.07$, $B=0.10$, $\gamma_{\rm max}=2.2\times 10^6$, 
$E_{\rm tot}=6.5\times 10^{11}$), ({\it T4}: $p=2.44$, $B=0.11$, 
$\gamma_{\rm max}=2.4\times 10^6$, $E_{\rm tot}=4.6\times 10^{11}$).}
\end{figure}

\begin{figure}
\epsscale{1.}
\plotone{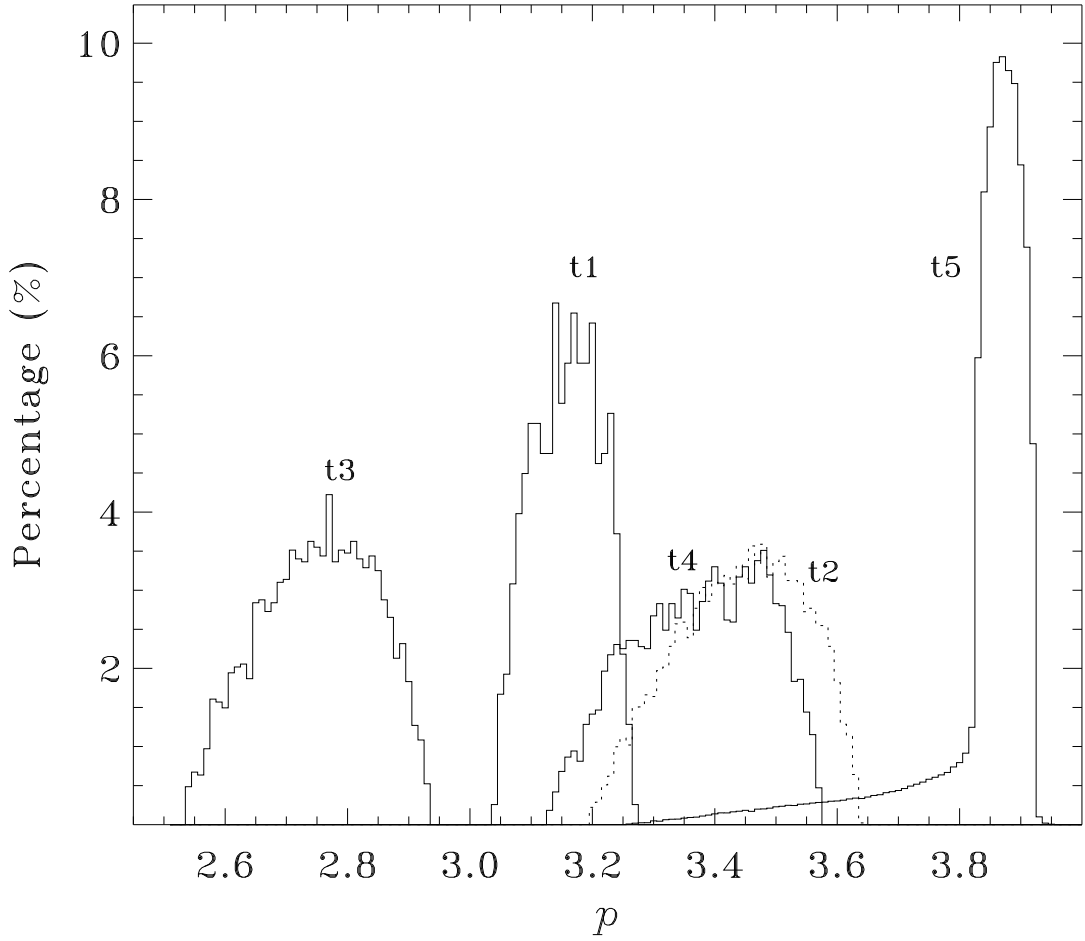}
\caption{Normalized distributions of the electron spectral index for Mrk~421. 
The results shown were derived for the case in which all key parameters are 
allowed to vary (see text). }
\end{figure}
\end{document}